\documentstyle[12pt]{article}
\font\titolo=cmbx12 scaled\magstep2

\font\tsnorm=cmr12

\font\tscorsp=cmti10

\def\NPB{Nucl. Phys. }

\def\PRD{Phys. Rev.  }

\def\CMP{ Comm. Math. Phys. }

\def\z{Z\kern -4.6pt Z}

\def\l{\lambda}

\def\z{\zeta}

\def\be{\begin{equation}}
\def\ee{\end{equation}}
\def\bea{\begin{eqnarray}}
\def\eea{\end{eqnarray}}
\def\bc{\begin{displaymath}}
\def\ec{\end{displaymath}}
\def\lb{\label}

\begin{document}
\pagestyle{empty}
\null
\vskip 5truemm
\begin{flushright}
INFNCA-TH9909\\
October 1999
\end{flushright}
\vskip 15truemm
\begin{center}
{\titolo Reply  Comment on ``Entropy of 2D black holes  }
\end{center}
\begin{center}
\titolo{from counting microstates''}
\end{center}
\vskip 15truemm
\begin{center}
{\tsnorm Mariano Cadoni$^{a,c,*}$ and Salvatore Mignemi$^{b,c,**}$}
\end{center}
%\smallskip
\begin{center}
{$^a$\tscorsp Dipartimento di Fisica,  
Universit\`a  di Cagliari,}
\end{center}
%\smallskip
\begin{center}
{\tscorsp Cittadella Universitaria, 09042, Monserrato, Italy.}
\end{center}
%\smallskip
%\smallskip
\begin{center}
{\tscorsp $^b$  Dipartimento di Matematica, Universit\'a  di Cagliari,}
\end{center}
%\smallskip
\begin{center}
{\tscorsp viale Merello 92, 09123, Cagliari, Italy.}
\end{center}
%\smallskip
\begin{center}
{\tscorsp $^c$  INFN, Sezione di Cagliari.}
\end{center}
\vskip 19truemm
%\baselineskip=2\normalbaselineskip
%%%%%%%%%%%%%%%%%%%%%%%%%%%%%%%%%%%%%%%%%%%%%%%%%%%%%%%%%%%%%%%%%%%%%%
%%		      abstract				       %%
%%%%%%%%%%%%%%%%%%%%%%%%%%%%%%%%%%%%%%%%%%%%%%%%%%%%%%%%%%%%%%%%%%%%%%
\begin{abstract}
\noindent
We show that the arguments proposed by Park and Yee against our recent 
derivation of the statistical entropy of 2D black holes do not apply 
to the case under consideration 
\end{abstract}
%%%%%%%%%%%%%%%%%%%%%%%%%%%%%%%%%%%%%%%%%%%%%%%%%%%%%%%%%%%%%%%%%%%%%%
%%			   End of abstract			       %%
%%%%%%%%%%%%%%%%%%%%%%%%%%%%%%%%%%%%%%%%%%%%%%%%%%%%%%%%%%%%%%%%%%%%%%
%%%%%%%%%%%%%%%%%%%%%%%%%%%%%%%%%%%%%%%%%%%%%%%%%%%%%%%%%%%%%%%%%%%%%%
%%			       Address				       %%
%%%%%%%%%%%%%%%%%%%%%%%%%%%%%%%%%%%%%%%%%%%%%%%%%%%%%%%%%%%%%%%%%%%%%%
\vfill
\begin{flushleft}
{\tsnorm PACS: 04.70. Dy, 04.50. +h  \hfill}
\end{flushleft}
\begin{flushleft}
\end{flushleft}
\smallskip
\vfill
\hrule
\begin{flushleft}
{$^*$E-Mail: CADONI@CA.INFN.IT\hfill}
\end{flushleft}
\begin{flushleft}
{$^{**}$E-Mail: MIGNEMI@CA.INFN.IT\hfill}
\end{flushleft}
\vfill
\eject
\pagenumbering{arabic}
\pagestyle{plain}

In a recent Comment \cite{PY} Park and Yee claimed that derivation of 
the statistical entropy of 2D (two-dimensional) black holes  proposed by 
us in Ref. 
\cite{CM} is plagued by an error.
In this Reply Comment  we show that the arguments used by  Park and Yee in 
Ref. \cite{PY}  against our derivation do not apply to our case.
Before going into the details of the confutation of the claim of Ref. 
\cite{PY}, let us briefly explain the arguments of Park and Yee.
  
In our attempt to calculate the statistical entropy of the 2D anti-de 
Sitter (AdS) black hole 
along the lines of Ref. \cite {ST} we found a major difficulty: owing 
to the dimension of the boundary, the  charges $J[\chi]$ 
(Eq. (18) of Ref. \cite{CM}) do not support a realization of the 
Virasoro algebra (the asymptotic symmetries of 2D AdS space). 
This problem is not a peculiarity 
of the 2D case but shows up also in higher dimensions \cite{ca}.
To solve the problem we proposed to define the new, time-integrated, 
generators $\hat J[\chi]$ (Eq. (22) of Ref. \cite{CM}). 
Moreover, we were able to show that the Dirac bracket algebra of the charges
$\hat J[\chi]$ gives a central extension of the Virasoro algebra and 
 to calculate its central charge.
To compute the central charge of the algebra we used the equation:
\be\lb{e1}
\widehat{\delta_{\omega}J[\chi]}=\hat J[[\chi,\omega]] + 
c(\chi,\omega),
\ee
where the hat has the meaning of an overall time-integration.
Park and Yee claim that the left-hand side of Eq. (\ref{e1}) can not be 
written as 
\be\lb{e2}
\{\hat J[\chi],\hat J[\omega]\}_{DB},
\ee 
thus invalidating our result that the charges $\hat J$ span a representation 
of the Virasoro algebra.  

The demonstration of Park and Yee 
relies on the two assumptions that  need 
to  be generalized if one  wants to interpret  consistently the
time-integrated charges $\hat J$ as  generators of a algebra.
First, the generators of the asymptotic 
symmetry can no longer  be identified with  the phase space functionals 
$H[\chi]$ ( Eq. (17) of 
Ref. \cite{CM}), but rather with the time-integrated ones $\hat H[\chi]$.
Second, the usual definition of the Poisson brackets,  as brackets 
evaluated at equal times,  has to be generalized in order to allow for 
general brackets
\be\lb{e3}
\{\hat H[\chi],\hat H[\omega]\}_{PB},
\ee
where $\hat H[\chi]$ is a time-integrated functional.
This  generalization of objects
 of the canonical formalism is implicitly contained in the definition 
 of the charges  $\hat J[\chi]$ \cite {CM1}. Moreover this is  what is
needed  in order to recognize Eq. (\ref{e1}) as  a canonical realization 
of the asymptotic symmetries. We do not know if in this framework the 
charges $\hat J[\chi]$ have a sensible interpretation as Noether 
charges. This is irrelevant for our purposes since we are just looking for 
a canonical realization of the asymptotic symmetries that allows us 
to perform  a computation of the central charge of the algebra.

Using the previously defined generalized notions of canonical 
generators and Poisson brackets we can easily  prove that the 
left-hand side of Eq. (\ref{e1}) can be written as a Dirac bracket 
algebra.  One just needs to compute explicitly the 
 brackets $\{\hat H[\chi],\hat H[\omega]\}_{PB}$. One finds \cite{CM1}:
\be\lb{e4}
\{\hat H[\chi],\hat H[\omega]\}=\hat H[[\chi,\omega]]+  c(\chi, \omega)
\ee
where the central charge has exactly the same value found in Ref. \cite {CM}. 
Fixing the gauge so that the constraints hold strongly and using Eq. (17)
of Ref. \cite{CM},
the previous equation implies
\be\lb{e5}
\{\hat J[\chi],\hat J[\omega]\}_{DB}=\hat J[[\chi,\omega]] + 
c(\chi,\omega).
\ee
Comparing this equation with Eq. (\ref{e1}), it follows immediately 
that  the left-hand side of the latter  can be written as the Dirac bracket 
in Eq. (\ref{e2}).
 
Let us now show explicitly that the calculations used in Ref. \cite {PY} by Park 
and Yee to  support   their claim are inconsistent with our 
definitions of  generators and Poisson brackets. From equation (\ref 
{e4}) it follows immediately that 
the canonical generators of the Virasoro algebra  are the functionals $\hat 
H[\chi]$ rather then  $H[\chi]$. Therefore Eq. (6) of  Ref. \cite{PY}, which is 
the starting point of the demonstration of Park and Yee,  does not 
apply.  The right equation to be used here is instead:

\be\lb{e6}
\{J[\chi],\hat H[\omega]\}_{DB}=\{J[\chi],\hat J[\omega]\}_{DB}=
{\delta_{\omega}J[\chi]}.
\ee

Following Yee and Park we perform now the time integration of Eq. 
(\ref{e6}).  The left-hand side becomes 
\be\lb{e7}
{\l\over 2\pi}\int^{2\pi\over \l}_{0}dt' \{J[\chi(t')],\hat 
J[\omega]\}_{DB}=\{\hat J[\chi],\hat J[\omega]\}_{DB},
\ee
from which it follows immediately that the left-hand side of Eq. 
(\ref{e1}) can be written as a Dirac bracket.

\end{document}